\documentclass[sigconf,nonacm]{acmart}
\usepackage{graphicx}
\usepackage{subfig}
\usepackage[autolanguage]{numprint}
\usepackage{listings}
\newcommand{\GHicon}{\raisebox{-.15\height}{\,\includegraphics[scale=0.02]{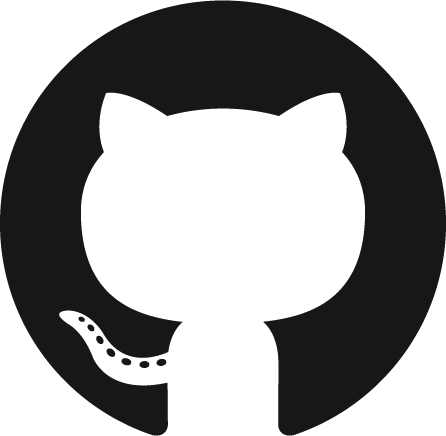} }}
\newcommand{\GAicon}{\raisebox{-.20\height}{\,\includegraphics[scale=0.07]{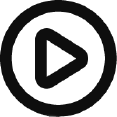} }}
\newcommand{\CMLicon}{\raisebox{-.15\height}{\,\includegraphics[scale=0.35]{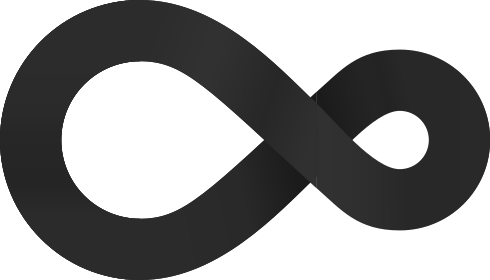} }}
\newcommand{\cml}{\textsc{CML}}
\newcommand{\ga}{\textsc{GitHub Actions}}

\AtBeginDocument{%
  \providecommand\BibTeX{{%
    \normalfont B\kern-0.5em{\scshape i\kern-0.25em b}\kern-0.8em\TeX}}}

\begin{document}

\title{A Preliminary Investigation of MLOps Practices in GitHub}

\author{Fabio Calefato}
\affiliation{%
  \institution{University of Bari}
  \streetaddress{Via Edoardo Orabona, 4}
  \city{Bari}
  \postcode{70125}
  \country{Italy}}
\email{fabio.calefato@uniba.it}
\orcid{0000-0003-2654-1588}

\author{Filippo Lanubile}
\affiliation{%
  \institution{University of Bari}
  \streetaddress{Via Edoardo Orabona, 4}
  \city{Bari}
  \postcode{70125}
  \country{Italy}}
\email{filippo.lanubile@uniba.it}
\orcid{0000-0003-3373-7589}

\author{Luigi Quaranta}
\affiliation{%
  \institution{University of Bari}
  \streetaddress{Via Edoardo Orabona, 4}
  \city{Bari}
  \postcode{70125}
  \country{Italy}}
\email{luigi.quaranta@uniba.it}
\orcid{0000-0002-9221-0739}

\renewcommand{\shortauthors}{Calefato et al.}

\begin{abstract}
  \textbf{Background}.~The rapid and growing popularity of machine learning (ML) applications has led to an increasing interest in MLOps, that is, the practice of continuous integration and deployment (CI/CD) of ML-enabled systems. 
  \textbf{Aims}.~Since changes may affect not only the code but also the ML model parameters and the data themselves, the automation of traditional CI/CD needs to be extended to manage model retraining in production.
  \textbf{Method}.~In this paper, we present an initial investigation of the MLOps practices implemented in a set of ML-enabled systems retrieved from GitHub, focusing on \ga\ and \cml, two solutions to automate the development workflow.
  \textbf{Results}.~Our preliminary results suggest that the adoption of MLOps workflows in open-source GitHub projects is currently rather limited.
  \textbf{Conclusions}.~Issues are also identified, which can guide future research work.
\end{abstract}


\begin{teaserfigure}
  \vspace{0.8cm}
\end{teaserfigure}

\maketitle

\section{Introduction}

ML-enabled systems, that is, software systems incorporating machine learning (ML) models, are receiving more and more attention from both researchers and practitioners~\citep{ozkaya_what_2020,lewis_characterizing_2021,nahar_collaboration_2021}.

Accordingly, there is also an increasing interest in MLOps~\citep{makinen_who_2021,treveil_introducing_2020}, that is, the development of solutions for the rapid delivery and deployment of ML models.
Because ML models are the core part of larger software systems, MLOps builds on DevOps and GitOps practices~\citep{ebert_devops_2016,limoncelli_gitops_2018}, and introduces additional actions that are specific to machine learning. 

Despite the increasing popularity of MLOps, studies on the adoption of automation and its impact on changes of ML-enabled systems are lacking in the academic literature.
As such, in this paper we aim to gather an initial understanding of how MLOps solutions are used to automate the execution of tasks to build and deploy ML-enabled systems, starting from open source projects available in GitHub.
In doing so, we make sure to exclude all ML-related projects that cannot be directly classified as ML-enabled systems or components: i.e., we discard all ML libraries, frameworks, and utilities, as they ultimately represent instances of traditional software projects for which the CI/CD practices have been already investigated (e.g., \citep{hilton_usage_2016}).

Then, we ask the following research questions:

\textbf{RQ1}. \textit{How common is workflow automation in ML-enabled systems hosted on GitHub?}

\textbf{RQ2}. \textit{What type of events are used to trigger MLOps workflows?}

\textbf{RQ3}. \textit{What are the most frequently enacted tasks?}

We answer these questions by means of an explorative archival study that mixes quantitative and qualitative analysis, focusing on \ga\ and \cml.
The former is a popular continuous integration and continuous delivery (CI/CD) platform, backed by GitHub, to automate software workflows,\footnote{\url{https://github.com/features/actions}} while the latter is a complementary solution to support continuous CI/CD practices that are specific to ML-enabled systems.\footnote{\url{https://cml.dev/doc}}

The remainder of the paper is structured as follows. 
In Sect.~\ref{sec:background}, we briefly illustrate how workflow automation works in \ga\ and \cml.
In Sect.~\ref{sec:study-design}, we present the design of the study, namely how we built the experimental datasets and the analyses conducted.
The results are presented in Sect.~\ref{sec:results} and their limitations in Sect.~\ref{sec:limitations}.
We discuss the findings and draw conclusions in Sect.~\ref{sec:discussion-conclusions}.

\section{Workflow Automation with \ga\ and \cml}\label{sec:background}

\ga\footnote{\url{https://docs.github.com/en/actions/learn-github-actions/understanding-github-actions}} is an event-driven API that automates development workflows in GitHub.
Workflows are defined by storing YAML files (e.g., \texttt{build.yml} or \texttt{deploy.yaml}) checked-in to a project's \texttt{.github/workflows/} directory, which will typically run when triggered by an \textit{event} occurring in the repository, such as a developer creating a pull request, opening an issue, or pushing a commit to a repository.

A workflow contains one or more jobs.
A \textit{job} is a set of steps that are executed in order on the same dedicated runner (typically, a container or a virtual machine) and are dependent on each other. 
Each step lets developers either execute a script or run an \textit{action}, a custom application for the \ga\ platform that performs a complex but frequently repeated task (e.g., set up the correct toolchain for a build environment, set up the cloud provider authentication, etc.).
As such, using actions helps reduce the amount of repetitive code written in  workflow files.
In addition, actions can be even published on the GitHub Marketplace for others to reuse. 

\begin{figure*}
    \centering
    \subfloat[Worfklow defintion in a YAML file]{\label{fig:ga-example-yml}
    \centering
    \includegraphics[width=0.25\linewidth]{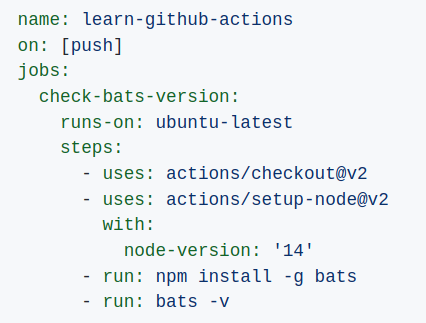}
    }
    \subfloat[Graphic representation of the same workflow]{\label{fig:ga-example-ui}
    \centering
    
    \includegraphics[width=0.40\linewidth]{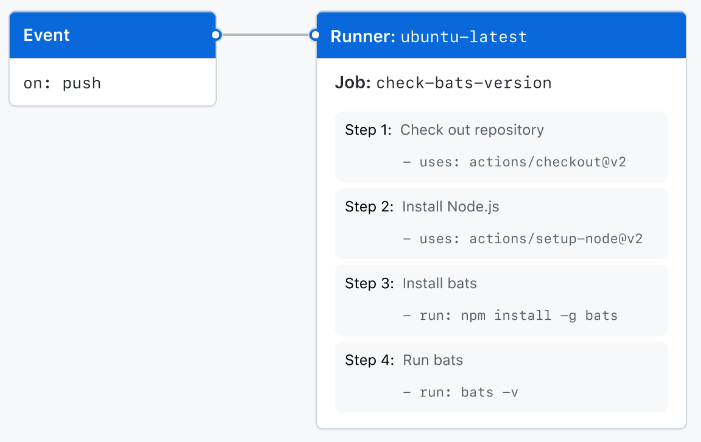}
    }
    \caption{An example of a \ga\ workflow.}
    \label{fig:ga-example}
\end{figure*}

Figure~\ref{fig:ga-example} shows an example of \ga\ as a YAML file (Figure~\ref{fig:ga-example-yml}) and as graphically displayed on GitHub (Figure~\ref{fig:ga-example-ui}).
The workflow is triggered on \texttt{push} events on the repository; it contains one job (i.e., \texttt{check-bats-version}) consisting of four consecutive steps, to be executed on an Ubuntu container as the chosen runner:
step~(1) and (2) use actions that, respectively, check out the source code and set up a NodeJS ver. 14 environment; then, step~(3) runs a custom command to install the \texttt{bats} library via the \texttt{npm} package manager; finally, step~(4) checks the version of the installed library.

\cml\footnote{\url{https://cml.dev}} is an open-source, command-line interface tool for implementing CI/CD in machine learning projects hosted on GitHub or GitLab.
In GitHub, \cml\ builds upon the \ga\ infrastructure, therefore the way it works is fairly  similar; in fact, developers must define jobs within a \texttt{cml.y*ml} file to be stored in the same \texttt{.github/workflows/} directory.
A typical \cml\ workflow can comprise shell commands, custom scripts, and specific \cml\ \textit{commands}.
\cml\ commands can be used, for instance, to commit changes to a new branch and open a Pull Request (PR), add comments to a PR, and start a runner.

In the rest of the paper, we use the term \textit{task} as an abstraction of Action in \ga\ and command in \cml.

\section{Study Design}\label{sec:study-design}

In this section, we describe the design of our study.
First, we describe the process adopted for building the two experimental datasets of MLOps workflows in GitHub, specific for \ga\ and \cml.
Then, we provide details about the data analysis method.

\begin{figure*}
    \centering
    \includegraphics[width=0.70\linewidth]{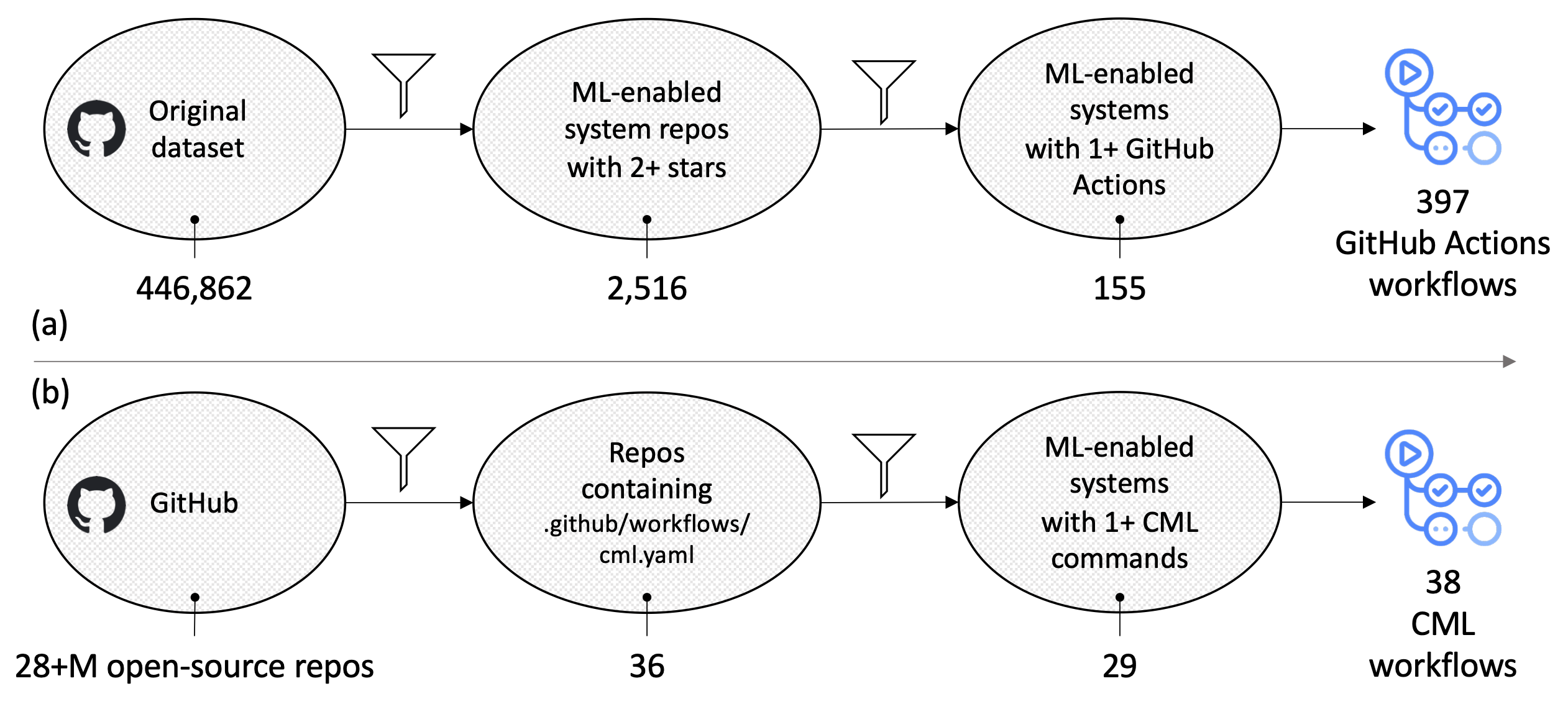}
    \caption{Steps applied to create the two datasets of \ga\ (a) and \cml\ (b) workflows.}
    \label{fig:dataset-creation}
\end{figure*}

\subsection{Dataset Construction}

\subsubsection{Identification of \ga\ Workflows}
The first step in our study was assembling an experimental dataset of \ga\ workflows taken from ML-related repositories.
We decided to focus on \ga\ as it is the default workflow automation tool offered by GitHub, the largest and the most popular code hosting platform for open-source projects.
Figure~\ref{fig:dataset-creation}a summarizes the steps of the dataset creation process.

We started from the dataset used in a recent work on the use of \ga~\cite{kinsman_how_2021},
comprising \numprint{446862} GitHub repositories.

In GitHub, users mark with a \textit{star} the repositories they like or are interested to; these become part of the list of their favorites, aka their \textit{stars}. In the GitHub jargon, users that mark a repository with a star are known as the repository \textit{stargazers}.
Many studies consider the number of stargazers to be evidence of repository popularity and -- as such -- a viable proxy of their quality \cite{munaiah_curating_2017, biswas_boa_2019}.
Hence, as a measure to keep low-quality projects out of our dataset, we decided to analyze only repositories having more than one star on GitHub (i.e., \numprint{183127} out of the \numprint{446862} repositories from the original set).

Moreover, \citeauthor{kalliamvakou_promises_2014} showed that a large portion of GitHub repositories are typically found to be inactive \cite{kalliamvakou_promises_2014}. In line with the goals of our study, we decided to consider only repositories that had been active for at least six months after the public release of GitHub Actions -- happened in November 2019. Specifically, we filtered out all repositories whose last-commit date preceded May 2020. In addition, to restrict our selection to ML-related repositories, we adopted a strategy similar to the one used by~\citeauthor{biswas_boa_2019} in~\citep{biswas_boa_2019}.
Specifically, we leveraged the GitHub API to retrieve the descriptions and topics of the repositories resulting from the filtering above.
Then, we sought ML systems-related keywords contained therein (e.g., \textit{`machine learning'}, \textit{`deep learning'}, \textit{`neural network'}, \textit{`image processing'}, etc.).\footnote{A complete list of the keywords we considered is available in the file \texttt{/settings.json} as part of the replication package of this study.}
Roughly, we used the same set of keywords adopted by~\citeauthor{biswas_boa_2019}, but also added a couple of noticeable absentees (i.e., \textit{`AI'} and \textit{`prediction model'}) and removed the subset of keywords referred to specific data science technologies (e.g., \texttt{`spark'}, \texttt{`hadoop'}, and \texttt{`cafe'}).
After applying the keyword-based filtering, we obtained a set of \numprint{2516} repositories.

For each candidate repository, we then checked the adoption of \ga\ by verifying the existence of YAML files (with extension \texttt{.yaml} or \texttt{.yml}) in the \texttt{./github/workflows} directory.
Out of the \numprint{2516} selected repositories, only 155 contain at least one \ga\ workflow.
Contextually, by leveraging the GitHub API, we downloaded all the available 399 workflows: almost all of them were valid YAML files, with only 2 exceptions.

In conclusion, our first dataset --- hereafter referred to as the \textit{\ga\ dataset} --- consists of 397 valid workflows extracted from 155 GitHub repositories.

\subsubsection{Identification of \cml\ Workflows}

To build the second dataset of projects containing CML workflows, we used the advanced search web interface to search globally across all of GitHub for repositories containing  \textit{.github/workflows/cml.y*ml} files.
We decided to focus on \cml\ because it is designed as a natural, ML-oriented extension of \ga\ and it is also tightly integrated with DVC,\footnote{\url{https://dvc.org}} one of the most popular data version control tools available to date~\citep{barrak_co-evolution_2021}.
Figure~\ref{fig:dataset-creation}b illustrates the steps for the dataset building process.

The query returned a list of only 36 candidates.
Given the few hits, in this case we did not apply any further filtering (e.g., number of stars as a proxy of quality).
Still, two authors manually vetted the list to exclude non-relevant results.
In particular, six repositories were discarded because they contain no ML models (2), no code (3), and an empty workflow file (1); we also discarded one repository that is merely the implementation of one of the tutorials available on the CML website.

The complete list of repositories is available as supplemental material in the replication package.\footnote{\url{https://github.com/collab-uniba/mlops4aisystems}}
We then used the GitHub API to download \ga\ workflows from the retained repositories.

In conclusion, our second dataset --- hereafter referred to as the \textit{\cml\ dataset} --- contains 41 \cml\ workflows extracted from 29 GitHub repositories, of which 38 valid.

\subsection{Data Analysis Method}

To answer RQ1 (i.e., \textit{``How common is workflow automation in ML-enabled systems hosted on GitHub?''}), we characterized the general adoption of \ga\ and \cml\ in our datasets by computing the number of repositories containing related workflows as well as the median number of workflows per repository. Then, two of the paper authors engaged in a manual inspection of the selected repositories to verify that they actually qualify as ML-enabled systems, discarding false positives from the filtering process.

To answer RQ2 (i.e., \textit{``What type of events are used to trigger MLOps workflows?''}), we computed the frequency of GitHub events used to trigger the workflows in our dataset.

Regarding RQ3 (i.e., \textit{``What are the most frequently enacted tasks?''}), to infer the most frequent tasks, we sought patterns among recurring Actions and shell commands from the collected workflows. 
Specifically, with the term \textit{`task'} here we refer to any logical unit of work that is typically found within an ML development workflow (e.g., `launching a model-retraining job,' `deploying an ML-enabled component to production,' or `saving a newly created model into a model registry'). 
Within \ga, such tasks can be implemented by applying one or more pre-defined actions, by running a sequence of shell commands, or even using a combination of both.
Therefore, to uncover patterns that may possibly constitute high-level tasks, we started by analyzing the most frequently used actions. 
Accordingly, we first computed the descriptive statistics of the actions found in our datasets, while also identifying the occurrence of those containing the words \texttt{`cml'} or `docker' in their slugs; indeed, we consider the use of these technologies a trace of typical MLOps practices (experiment tracking and ML models deployment, respectively). 
Then, we leveraged the \textit{Apriori} algorithm to compute the set of frequently co-occurring actions;
to do so, we grouped action slugs (stripped of their version tags) by the workflows in which they appear, thus generating a collection of transactions.\footnote{
In frequent pattern mining, a transaction is defined as a collection of items that have been observed together; in our case, each transaction corresponds to the collection of actions observed together in the context of a single workflow.}

As in \ga\ the steps of a workflow job may consist of either actions (denoted by the attribute \texttt{uses}) or shell commands/scripts (denoted by \texttt{run}), we applied a similar analytical procedure -- including the application of the \textit{Apriori} algorithm -- to particular subsets of shell commands from the \texttt{run} steps. 
Specifically, we studied the (co-)occurrence of shell commands containing the keywords \texttt{cml} and \texttt{docker}.

Finally, we performed a qualitative analysis of the workflows by manually inspecting the most interesting ones, i.e., those belonging to repositories that contain ML-enabled software components.

\section{Results}\label{sec:results}

In this section, we present and discuss the results of our preliminary investigation of MLOps practices in GitHub.

\subsection{RQ1.  How common is workflow automation in ML-enabled systems hosted on GitHub?}\label{sec:res-rq1}

\subsubsection{Results from the \ga\ dataset}

Our \ga\ dataset comprises 397 valid GitHub Actions workflows distributed across 155 repositories. The median number of workflows contained in each repository is 2, while the third quartile of the distribution is 3 and the maximum number of workflows found in a repository is 14.

Having retrieved only a limited number of repositories, we could verify the results of the filtering process with a manual inspection. 
Two authors inspected each repository on GitHub and checked that it actually contained an ML-enabled system or an ML-enabled software component to be integrated in a larger system. 
The vast majority of the repositories (105) have been misclassified as ML-related projects; the remaining repositories (50) are ML libraries or frameworks (e.g., \GHicon \texttt{\href{https://github.com/dmlc/xgboost}{dmlc/xgboost}}, \GHicon \texttt{\href{https://github.com/mlpack/mlpack}{mlpack/mlpack}},  \GHicon \texttt{\href{https://github.com/h2oai/h2o-dev}{h2oai/h2o-dev}}), mostly written in Python or Java.
We decided to discard these kinds of repositories as -- apart from their ML-oriented goal -- they actually look no different from traditional software projects, with which they share the typical engineering requirements and needs (e.g., the use of traditional unit testing, the adoption of code quality assurance tools, etc.)\citep{ozkaya_what_2020}.
By looking at the misclassified instances, we noticed that the `\textit{machine learning}' and `\textit{artificial intelligence}' topic labels are often used as buzzwords due to the current hype.

Only one repository can be considered an actual example of ML-enabled system (see \GHicon \texttt{\href{https://github.com/tesseract-ocr/tesseract}{tesseract-ocr/tesseract}}). 
However, the related \ga\ workflows are devoted to testing and benchmarking the application as a whole and none of them seems to address any ML task directly.

Given the young age of mature ML techniques, we speculate that most of the professional collaboration around ML-enabled components and systems might currently be happening within companies and private research institutions. 
In such contexts, high-performance ML models constitute a valuable economical asset and sharing them as open-source software would likely spoil their value. 
Consequently, we argue that ML-enabled systems may be more frequently found in private GitHub repositories rather then in public ones.

Having not found relevant material for our study in the \ga\ dataset, after our manual inspection, we decided to discard it. 
As such, the analytical results hereafter presented are exclusively referred to the \cml\ dataset discussed next.

\subsubsection{Results from the \cml\ dataset}

The \cml\ dataset comprises 38 valid GitHub Actions workflows distributed across 29 repositories.
The median number of workflows contained in each repository is 1 and the third quartile of the distribution is still 1; the maximum number of workflows found in a repository is 4.

Also in this case, two authors manually vetted the dataset, individually examining the contents of the retrieved repositories. As a result of this manual inspection, we classified most of the repositories from this collection (24) as \textit{test-driving} (i.e., repositories containing proof-of-concept ML projects, likely used by GitHub users to test-drive the \cml\ library). A couple of repositories are explicitly linked to educational material, i.e., a university ML course (\GHicon \texttt{\href{https://github.com/tue-5ARA0/mlops-demo-live}{tue-5ARA0/mlops-demo-live}}) and a technical blog post (\GHicon \texttt{\href{https://github.com/amitvkulkarni/Bring-DevOps-to-Machine-Learning-with-CML}{amitvkulkarni/Bring-DevOps-to-Machine-Learning-\\with-CML}}); as such, we classify them as \textit{educational.} 
Finally, in 3 repositories we found small-size ML projects (i.e., up to three contributors), to which we refer as \textit{ML component}:
the first (\GHicon \texttt{\href{https://github.com/AscendNTNU/perception_testing_21}{Ascend\\NTNU/perception\_testing\_21}}) is a workspace for building models for ROS, a robot operating system;
the second (\GHicon \texttt{\href{https://github.com/cheesama/morphine}{cheesama/mor\\phine}}) an entity classifier in Korean for the pynori Python library;
the third (\GHicon \texttt{\href{https://github.com/mozartofmath/AmharicSpeechToText}{mozartofmath/AmharicSpeechToText}}) a speech-to-text engine for the Amharic language.

Also in this case, we observe a substantial lack of production-grade ML projects containing traces of MLOps practices. 
However, all repositories from this dataset contain \cml\ workflows, each of which is specifically defined to accomplish one or more MLOps tasks. 
As such, despite many of them having testing or educational purposes, we deemed all repositories from this dataset worth of being analyzed.

\subsection{RQ2. What type of events are used to trigger MLOps workflows?}

Most of the workflows from our dataset are triggered by \texttt{git} \texttt{push} and \texttt{pull\_request} events: they are used in 78.95\% and 18.42\% of the workflows, respectively. 
Moreover, these events co-occur as triggers in 13.16\% of the workflows. 
Instead, GitHub-managed events are rarely used to activate workflows.
Only 3 of them start as a consequence of a new \texttt{issue\_comment} and just 2 are triggered by GitHub's \texttt{release} event.
The \texttt{schedule} event -- allowing workflows activation at specific times -- is also used only twice.

While it is not surprising that the native \texttt{git} events \texttt{push} and \texttt{pull\_request} are the most commonly used triggers for projects hosted on GitHub, we would have expected a prevalence of workflows activated by \texttt{pull\_request} and \texttt{release} events.
Indeed, typical MLOps tasks are computationally heavy and time consuming (e.g., model re-training and testing) and, as such, presumably expansive to run every time a new push operation is performed.

\subsection{RQ3. What are the most frequently enacted tasks?}

To identify potential ML-related tasks accomplished within the collected workflows, we analyzed the pre-defined actions used in each of them.
The total number of distinct actions retrieved in the \cml\ dataset is 15 (14 if we ignore version tags). 
Typically, actions are used slightly more than run commands: the average number of workflow steps using actions over the total number of workflow steps is 0.53.
The median number of actions in each workflow is 3 (mean: 2.55, standard deviation: 1.20); no workflow uses more than 5 actions.
Of the 14 distinct actions adopted by the workflows in our dataset, 11 (\textasciitilde 79\%) are available in the GitHub Marketplace and 8 of them are by creators verified by GitHub.\footnote{In the GitHub Marketplace, Actions get this label when they are developed by creators that have an established relationship with GitHub; these Actions are typically the result of the joint work of GitHub and their creators.}
Moreover, we were able to find only one action containing the substring `\textit{cml}' in its slug (i.e., \texttt{setup-cml}), while no slug contained the word `\textit{docker}'.

By using the \textit{Apriori} algorithm, we computed the frequent sets of actions with support higher than 0.07 (i.e., appearing in at least 3 workflows); the results are reported in Table~\ref{tab:co-occurring-actions}.
The top-4 frequent itemsets combine \texttt{actions/checkout} with \texttt{actions/setup-cml} and \texttt{actions/setup-python}.
Actually, \texttt{actions/checkout} is found in almost every frequent itemset, as it is commonly required in all jobs aimed at processing the contents of a repository.

\begin{table*}[t]
    \caption{Frequently co-occurring actions with support > 0.07 (i.e., actions co-occurring in at least 3 workflows).}
    \centering
    \begin{tabular}{lllr}
    \toprule
                \GAicon \textbf{Action~1} &                        \GAicon \textbf{Action~2} &                \GAicon \textbf{Action~3} & \textbf{Support} \\
    \midrule
        actions/checkout &             iterative/setup-cml &                         &    0.47 (18 workflows) \\
        actions/checkout &            actions/setup-python &                         &    0.37 (14 workflows) \\
        actions/setup-cml &           actions/setup-python &                         &    0.34 (13 workflows) \\
        actions/checkout &            actions/setup-python &                        actions/setup-cml &    0.34 (13 workflows) \\
        actions/checkout &  aws-actions/configure-aws-credentials &
             &    0.11 (4 workflows) \\
        actions/checkout &          actions/setup-go &                         &    0.08 (3 workflows) \\
    \bottomrule
    \end{tabular}
    \vspace{-5pt}
    \label{tab:co-occurring-actions}
\end{table*}

If we limit our analysis to the workflows containing `\textit{cml}' in their filename, we find that only 6 distinct actions are used. Table~\ref{tab:actions-in-cml-workflows} presents each of them together with their frequencies. 
In the following, we briefly report their definition:

\begin{description}
  \item[
      \GAicon\href{https://github.com/marketplace/actions/checkout}
      {\texttt{actions/checkout}}
  ]
  Checks-out the contents of a repository in the virtual environment in which a job is running;
  
  \item[
    \GAicon\href{https://github.com/iterative/setup-cml}
    {\texttt{iterative/setup-cml}}
  ]
  Enables CML functionalities with-  in \ga\ workflows.
  
  \item[
    \GAicon\href{https://github.com/marketplace/actions/setup-python}
    {\texttt{actions/setup-python}}
  ]
  Sets up the requested version of Python, to be used in the subsequent steps of a job.
  
  \item[
    \GAicon\href{https://github.com/iterative/setup-dvc}
    {\texttt{iterative/setup-dvc}}
  ]
  Sets up DVC (Data Version Control) functionalities within \ga\ workflows.
  
  \item[
    \GAicon\href{https://github.com/ros-industrial/industrial_ci}
    {\texttt{ros-industrial/industrial\_ci}}
  ]
  Configures a CI process specifically tailored for packages powered by ROS (Robot Operating System).
  
  \item[
    \GAicon\href{https://github.com/marketplace/actions/configure-aws-credentials-action-for-github-actions}
    {\texttt{aws-actions/configure-aws-credentials}}
  ]
  Configures AWS credentials and region environment variables to be used for AWS API calls.
\end{description}

\begin{table}[]
    \caption{Actions found within \cml\ workflows.}
    \centering
    \begin{tabular}{lr}
    \toprule
    \GAicon \textbf{Action} &  \textbf{Frequency} \\
    \midrule
    actions/checkout                        &                28 \\
    iterative/setup-cml                     &                17 \\
    actions/setup-python                    &                13 \\
    iterative/setup-dvc                     &                2 \\
    ros-industrial/industrial\_ci           &                1 \\
    aws-actions/configure-aws-credentials   &                1 \\
    \bottomrule
    \end{tabular}
    \label{tab:actions-in-cml-workflows}
\end{table}

Every \cml\ workflow uses \texttt{actions/checkout} to make the contents of the repository available in the Actions
runner; then, about half of them set up \cml\ and Python using pre-defined actions. 
Noticeably, \GAicon\texttt{iterative/setup-cml} is not an action verified by GitHub and is not yet available in the Marketplace.
We observe that none of these actions can be used alone to perform an high-level MLOps task. 
Instead, they are primarily used to set up the CI environment and accomplish preliminary configuration steps.

After characterizing the use of actions in our dataset, we delved into the analyses of shell commands extracted from \texttt{run} attributes.
Of the 38 \ga\ workflows from the \cml\ dataset, 28 present shell commands containing `\textit{cml}' as a substring. \cml\ offers a few commands to support the implementation of CI/CD in ML projects; here we report a brief definition of those found in our workflows:

\begin{description}
  \item[
      \CMLicon\href{https://cml.dev/doc/ref/send-comment}
      {\texttt{send-comment}}
  ]
  adds a Markdown report as a comment on a commit or pull/merge request.
  
  \item[
    \CMLicon\href{https://cml.dev/doc/ref/publish}
    {\texttt{publish}}
  ]
  uploads and publicly hosts an image (e.g., a \texttt{.png} plot) for displaying in a CML report.
  
  \item[
    \CMLicon\href{https://cml.dev/doc/ref/runner}
    {\texttt{runner}}
  ]
  starts a runner in any supported cloud compute provider (or locally, in case of an on-premise \cml\ deployment).
  
  \item[
    \CMLicon\href{https://cml.dev/doc/ref/tensorboard-dev}
    {\texttt{tensorboard-dev}}
  ]
  returns a link to \href{https://tensorboard.dev}{tensorboard.dev}, a managed TensorBoard\footnote{\href{https://www.tensorflow.org/tensorboard?hl=en}{TensorBoard} is a visualization tool for tracking ML experiments. It is typically used in combination with \href{https://www.tensorflow.org/?hl=en}{TensorFlow}, a popular open-source experimentation platform for ML.} platform that lets users upload and share their ML experiment results.
  
\end{description}

All 28 workflows containing a \cml\ command use \CMLicon\texttt{send-comm\\ent} to decorate a commit or pull-request message with model training and evaluation metrics. In 17 cases (\textasciitilde 63\%), the \texttt{publish} command is used along with \CMLicon\texttt{send-comment} to add an image to the Markdown-formatted message. Only in 2 distinct cases we observed the use of \CMLicon\texttt{runner} and \CMLicon\texttt{tensorboard-dev}. As the \CMLicon\texttt{send-comment} and \CMLicon\texttt{publish} are used just for reporting purposes, we find particularly interesting the adoption of \CMLicon\texttt{runner} in (\GHicon \texttt{\href{https://github.com/ibrahimkaratas88/cml_cloud_try}{ibrahimkaratas88/cml\_cloud\_try}}); by manually inspecting the workflow, we found that it is used to set up an AWS virtual machine to be used as a self-hosted runner in a subsequent job. In the following, the workflow uses DVC to reproduce  the data-transfer task on the cloud machine and report the related results via the \CMLicon\texttt{send-comment} command. 
However, as far as we can tell, model training or re-training tasks are not involved in this example.

In search for end-to-end MLOps pipelines, we also sought the use of Docker commands. 
Within our dataset, Docker is used just once (in \GHicon \texttt{\href{https://github.com/2796gaurav/automate}{2796gaurav/automate}}, classified as a \textit{test-driving} repository). 
Specifically, the \texttt{build} and \texttt{push} Docker commands are used in \texttt{dockerize.yml} to build and upload a Docker image to AWS Elastic Container Registry using Terraform, a multi-cloud tool to implement infrastructure as code.
This workflow is executed only when a message containing the string \textit{`/dockerize'} is published as a comment in an issue or pull request.

Along with this and a few other \ga\ workflows, the same repository contains a \cml\ workflow, named \texttt{cml\_report.yml}, which uses DVC to reproduce a full ML pipeline (from data preparation to model evaluation) and then makes training results available on GitHub via the \GAicon\texttt{cml-send-comment} command.

\section{Limitations}\label{sec:limitations}
In this section, we illustrate the main limitations of our study and our plans to overcome them.
For the sake of transparency and repeatability, we have made our scripts and data publicly available.

Being based on the analysis of a very limited sample of ML-enabled projects, in its current form, our study lacks external validity. 
To overcome this, we plan on expanding our work along different paths.
Upon inspecting the sample of open-source GitHub repositories from the dataset used in \cite{kinsman_how_2021}, we could not find any relevant project for our study. 
Nevertheless, we believe that other datasets assembled from GitHub might be worth investigating as well: for instance, we are currently analyzing the corpus of ML projects gathered by \citeauthor{gonzalez_state_2020} \cite {gonzalez_state_2020}. 
Moreover, we intend to take into account additional code hosting platforms, such as GitLab and Bitbucket, as they might also represent a valuable source for ML-enabled systems.
Furthermore, as already discussed in Sect.~\ref{sec:res-rq1}, we argue that many interesting repositories containing ML-enabled components or systems might not be currently available as open-source software. 
To take into account the adoption of MLOps practices in the industry, we are now designing a couple of case studies to be investigated at specialized companies. 

Another limitation of our study is the focus on two particular software solutions (i.e., \ga\ and \cml). The landscape of CI/CD tools and MLOps platforms is large and constantly evolving. 
In a future study, we plan to take into consideration further popular CI/CD solutions, such as GitLab CI/CD, TravisCI, and CircleCI. 
Also, we will take into account alternatives to \cml\ as a specialized MLOps solution: indeed, despite having been publicly available for about one year now, \cml\ appears to be still scarcely adopted in GitHub and mostly for test-driving purposes.

Finally, in our study, we considered the use of Docker a viable proxy to investigate deployment practices for ML-enabled components. 
In our future work, we plan on considering additional deployment strategies and tools, e.g., by studying deployment solutions offered by commercial cloud providers and their integration into MLOps workflows.

\section{Conclusions}\label{sec:discussion-conclusions}

In this paper, we present a preliminary investigation of MLOps practices in GitHub. Overall --- despite the popularity of the code hosting platform --- we highlight a substantial lack of open-source projects concerning ML-enabled systems or components which leverage \ga; arguably, it is possible that such kind of projects might be hosted as private repositories. 

Conversely, many ML-related open-source libraries and frameworks leverage \ga.
However, these cannot be deemed relevant for the study of MLOps practices as they are not different from traditional software systems, apart from the ML-related goal or use-case.

Searching for projects that leverage \cml -- a specialized MLOps tool -- also yielded a few results.
Therefore, researchers interested in studying MLOps practices might want to consider broadening the scope of potential sources by taking into account further code hosting platforms (e.g., GitLab and Bitbucket), CI/CD services (e.g., TravisCI and CircleCI), and specialized MLOps tools.

\bibliographystyle{ACM-Reference-Format}
\bibliography{references}


\begin{thebibliography}{14}


\ifx \showCODEN    \undefined \def \showCODEN     #1{\unskip}     \fi
\ifx \showDOI      \undefined \def \showDOI       #1{#1}\fi
\ifx \showISBNx    \undefined \def \showISBNx     #1{\unskip}     \fi
\ifx \showISBNxiii \undefined \def \showISBNxiii  #1{\unskip}     \fi
\ifx \showISSN     \undefined \def \showISSN      #1{\unskip}     \fi
\ifx \showLCCN     \undefined \def \showLCCN      #1{\unskip}     \fi
\ifx \shownote     \undefined \def \shownote      #1{#1}          \fi
\ifx \showarticletitle \undefined \def \showarticletitle #1{#1}   \fi
\ifx \showURL      \undefined \def \showURL       {\relax}        \fi
\providecommand\bibfield[2]{#2}
\providecommand\bibinfo[2]{#2}
\providecommand\natexlab[1]{#1}
\providecommand\showeprint[2][]{arXiv:#2}

\bibitem[Barrak et~al\mbox{.}(2021)]%
        {barrak_co-evolution_2021}
\bibfield{author}{\bibinfo{person}{Amine Barrak}, \bibinfo{person}{Ellis~E.
  Eghan}, {and} \bibinfo{person}{Bram Adams}.} \bibinfo{year}{2021}\natexlab{}.
\newblock \showarticletitle{On the {Co}-evolution of {ML} {Pipelines} and
  {Source} {Code} - {Empirical} {Study} of {DVC} {Projects}}. In
  \bibinfo{booktitle}{\emph{2021 {IEEE} {International} {Conference} on
  {Software} {Analysis}, {Evolution} and {Reengineering} ({SANER})}}.
  \bibinfo{publisher}{IEEE}, \bibinfo{address}{Honolulu, HI, USA},
  \bibinfo{pages}{422--433}.
\newblock
\showISBNx{978-1-72819-630-5}
\urldef\tempurl%
\url{https://doi.org/10.1109/SANER50967.2021.00046}
\showDOI{\tempurl}


\bibitem[Biswas et~al\mbox{.}(2019)]%
        {biswas_boa_2019}
\bibfield{author}{\bibinfo{person}{Sumon Biswas}, \bibinfo{person}{Md~Johirul
  Islam}, \bibinfo{person}{Yijia Huang}, {and} \bibinfo{person}{Hridesh
  Rajan}.} \bibinfo{year}{2019}\natexlab{}.
\newblock \showarticletitle{Boa {Meets} {Python}: {A} {Boa} {Dataset} of {Data}
  {Science} {Software} in {Python} {Language}}. In
  \bibinfo{booktitle}{\emph{2019 {IEEE}/{ACM} 16th {International} {Conference}
  on {Mining} {Software} {Repositories} ({MSR})}}. \bibinfo{publisher}{IEEE},
  \bibinfo{address}{Montreal, QC, Canada}, \bibinfo{pages}{577--581}.
\newblock
\showISBNx{978-1-72813-412-3}
\urldef\tempurl%
\url{https://doi.org/10.1109/MSR.2019.00086}
\showDOI{\tempurl}


\bibitem[Ebert et~al\mbox{.}(2016)]%
        {ebert_devops_2016}
\bibfield{author}{\bibinfo{person}{Christof Ebert}, \bibinfo{person}{Gorka
  Gallardo}, \bibinfo{person}{Josune Hernantes}, {and} \bibinfo{person}{Nicolas
  Serrano}.} \bibinfo{year}{2016}\natexlab{}.
\newblock \showarticletitle{{DevOps}}.
\newblock \bibinfo{journal}{\emph{Ieee Software}} \bibinfo{volume}{33},
  \bibinfo{number}{3} (\bibinfo{year}{2016}), \bibinfo{pages}{94--100}.
\newblock
\newblock
\shownote{Publisher: IEEE}.


\bibitem[Gonzalez et~al\mbox{.}(2020)]%
        {gonzalez_state_2020}
\bibfield{author}{\bibinfo{person}{Danielle Gonzalez}, \bibinfo{person}{Thomas
  Zimmermann}, {and} \bibinfo{person}{Nachiappan Nagappan}.}
  \bibinfo{year}{2020}\natexlab{}.
\newblock \showarticletitle{The {State} of the {ML}-universe: 10 {Years} of
  {Artificial} {Intelligence} \& {Machine} {Learning} {Software} {Development}
  on {GitHub}}. In \bibinfo{booktitle}{\emph{Proceedings of the 17th
  {International} {Conference} on {Mining} {Software} {Repositories}}}.
  \bibinfo{publisher}{ACM}, \bibinfo{address}{Seoul Republic of Korea},
  \bibinfo{pages}{431--442}.
\newblock
\showISBNx{978-1-4503-7517-7}
\urldef\tempurl%
\url{https://doi.org/10.1145/3379597.3387473}
\showDOI{\tempurl}


\bibitem[Hilton et~al\mbox{.}(2016)]%
        {hilton_usage_2016}
\bibfield{author}{\bibinfo{person}{Michael Hilton}, \bibinfo{person}{Timothy
  Tunnell}, \bibinfo{person}{Kai Huang}, \bibinfo{person}{Darko Marinov}, {and}
  \bibinfo{person}{Danny Dig}.} \bibinfo{year}{2016}\natexlab{}.
\newblock \showarticletitle{Usage, costs, and benefits of continuous
  integration in open-source projects}. In
  \bibinfo{booktitle}{\emph{Proceedings of the 31st {IEEE}/{ACM}
  {International} {Conference} on {Automated} {Software} {Engineering}}}.
  \bibinfo{publisher}{ACM}, \bibinfo{address}{Singapore Singapore},
  \bibinfo{pages}{426--437}.
\newblock
\showISBNx{978-1-4503-3845-5}
\urldef\tempurl%
\url{https://doi.org/10.1145/2970276.2970358}
\showDOI{\tempurl}


\bibitem[Kalliamvakou et~al\mbox{.}(2014)]%
        {kalliamvakou_promises_2014}
\bibfield{author}{\bibinfo{person}{Eirini Kalliamvakou},
  \bibinfo{person}{Georgios Gousios}, \bibinfo{person}{Kelly Blincoe},
  \bibinfo{person}{Leif Singer}, \bibinfo{person}{Daniel~M. German}, {and}
  \bibinfo{person}{Daniela Damian}.} \bibinfo{year}{2014}\natexlab{}.
\newblock \showarticletitle{The promises and perils of mining {GitHub}}. In
  \bibinfo{booktitle}{\emph{Proceedings of the 11th {Working} {Conference} on
  {Mining} {Software} {Repositories} - {MSR} 2014}}. \bibinfo{publisher}{ACM
  Press}, \bibinfo{address}{Hyderabad, India}, \bibinfo{pages}{92--101}.
\newblock
\showISBNx{978-1-4503-2863-0}
\urldef\tempurl%
\url{https://doi.org/10.1145/2597073.2597074}
\showDOI{\tempurl}


\bibitem[Kinsman et~al\mbox{.}(2021)]%
        {kinsman_how_2021}
\bibfield{author}{\bibinfo{person}{Timothy Kinsman}, \bibinfo{person}{Mairieli
  Wessel}, \bibinfo{person}{Marco~A. Gerosa}, {and} \bibinfo{person}{Christoph
  Treude}.} \bibinfo{year}{2021}\natexlab{}.
\newblock \showarticletitle{How {Do} {Software} {Developers} {Use} {GitHub}
  {Actions} to {Automate} {Their} {Workflows}?}. In
  \bibinfo{booktitle}{\emph{2021 {IEEE}/{ACM} 18th {International} {Conference}
  on {Mining} {Software} {Repositories} ({MSR})}}. \bibinfo{publisher}{IEEE},
  \bibinfo{address}{Madrid, Spain}, \bibinfo{pages}{420--431}.
\newblock
\showISBNx{978-1-72818-710-5}
\urldef\tempurl%
\url{https://doi.org/10.1109/MSR52588.2021.00054}
\showDOI{\tempurl}


\bibitem[Lewis et~al\mbox{.}(2021)]%
        {lewis_characterizing_2021}
\bibfield{author}{\bibinfo{person}{Grace~A. Lewis}, \bibinfo{person}{Stephany
  Bellomo}, {and} \bibinfo{person}{Ipek Ozkaya}.}
  \bibinfo{year}{2021}\natexlab{}.
\newblock \showarticletitle{Characterizing and {Detecting} {Mismatch} in
  {Machine}-{Learning}-{Enabled} {Systems}}. In \bibinfo{booktitle}{\emph{2021
  {IEEE}/{ACM} 1st {Workshop} on {AI} {Engineering} - {Software} {Engineering}
  for {AI} ({WAIN})}}. \bibinfo{publisher}{IEEE}, \bibinfo{address}{Madrid,
  Spain}.
\newblock
\urldef\tempurl%
\url{https://doi.org/10.1109/WAIN52551.2021.00028}
\showDOI{\tempurl}


\bibitem[Limoncelli(2018)]%
        {limoncelli_gitops_2018}
\bibfield{author}{\bibinfo{person}{Thomas~A. Limoncelli}.}
  \bibinfo{year}{2018}\natexlab{}.
\newblock \showarticletitle{{GitOps}: {A} {Path} to {More} {Self}-{Service}
  {IT}: {IaC} + {PR} = {GitOps}}.
\newblock \bibinfo{journal}{\emph{Queue}} \bibinfo{volume}{16},
  \bibinfo{number}{3} (\bibinfo{date}{June} \bibinfo{year}{2018}),
  \bibinfo{pages}{13--26}.
\newblock
\showISSN{1542-7730}
\urldef\tempurl%
\url{https://doi.org/10.1145/3236386.3237207}
\showDOI{\tempurl}
\newblock
\shownote{Place: New York, NY, USA Publisher: Association for Computing
  Machinery}.


\bibitem[Makinen et~al\mbox{.}(2021)]%
        {makinen_who_2021}
\bibfield{author}{\bibinfo{person}{Sasu Makinen}, \bibinfo{person}{Henrik
  Skogstrom}, \bibinfo{person}{Eero Laaksonen}, {and} \bibinfo{person}{Tommi
  Mikkonen}.} \bibinfo{year}{2021}\natexlab{}.
\newblock \showarticletitle{Who {Needs} {MLOps}: {What} {Data} {Scientists}
  {Seek} to {Accomplish} and {How} {Can} {MLOps} {Help}?}. In
  \bibinfo{booktitle}{\emph{2021 {IEEE}/{ACM} 1st {Workshop} on {AI}
  {Engineering} - {Software} {Engineering} for {AI} ({WAIN})}}.
  \bibinfo{publisher}{IEEE}, \bibinfo{address}{Madrid, Spain},
  \bibinfo{pages}{109--112}.
\newblock
\showISBNx{978-1-66544-470-5}
\urldef\tempurl%
\url{https://doi.org/10.1109/WAIN52551.2021.00024}
\showDOI{\tempurl}


\bibitem[Munaiah et~al\mbox{.}(2017)]%
        {munaiah_curating_2017}
\bibfield{author}{\bibinfo{person}{Nuthan Munaiah}, \bibinfo{person}{Steven
  Kroh}, \bibinfo{person}{Craig Cabrey}, {and} \bibinfo{person}{Meiyappan
  Nagappan}.} \bibinfo{year}{2017}\natexlab{}.
\newblock \showarticletitle{Curating {GitHub} for engineered software
  projects}.
\newblock \bibinfo{journal}{\emph{Empirical Software Engineering}}
  \bibinfo{volume}{22}, \bibinfo{number}{6} (\bibinfo{date}{Dec.}
  \bibinfo{year}{2017}), \bibinfo{pages}{3219--3253}.
\newblock
\showISSN{1382-3256, 1573-7616}
\urldef\tempurl%
\url{https://doi.org/10.1007/s10664-017-9512-6}
\showDOI{\tempurl}


\bibitem[Nahar et~al\mbox{.}(2021)]%
        {nahar_collaboration_2021}
\bibfield{author}{\bibinfo{person}{Nadia Nahar}, \bibinfo{person}{Shurui Zhou},
  \bibinfo{person}{Grace Lewis}, {and} \bibinfo{person}{Christian Kästner}.}
  \bibinfo{year}{2021}\natexlab{}.
\newblock \showarticletitle{Collaboration {Challenges} in {Building}
  {ML}-{Enabled} {Systems}: {Communication}, {Documentation}, {Engineering},
  and {Process}}.
\newblock \bibinfo{journal}{\emph{arXiv:2110.10234 [cs]}} (\bibinfo{date}{Dec.}
  \bibinfo{year}{2021}).
\newblock
\urldef\tempurl%
\url{http://arxiv.org/abs/2110.10234}
\showURL{%
\tempurl}
\newblock
\shownote{arXiv: 2110.10234}.


\bibitem[Ozkaya(2020)]%
        {ozkaya_what_2020}
\bibfield{author}{\bibinfo{person}{Ipek Ozkaya}.}
  \bibinfo{year}{2020}\natexlab{}.
\newblock \showarticletitle{What is really different in engineering ai-enabled
  systems?}
\newblock \bibinfo{journal}{\emph{IEEE Software}} \bibinfo{volume}{37},
  \bibinfo{number}{4} (\bibinfo{year}{2020}), \bibinfo{pages}{3--6}.
\newblock
\newblock
\shownote{Publisher: IEEE}.


\bibitem[Treveil et~al\mbox{.}(2020)]%
        {treveil_introducing_2020}
\bibfield{author}{\bibinfo{person}{Mark Treveil}, \bibinfo{person}{Nicolas
  Omont}, \bibinfo{person}{Clément Stenac}, \bibinfo{person}{Kenji Lefevre},
  \bibinfo{person}{Du Phan}, \bibinfo{person}{Joachim Zentici},
  \bibinfo{person}{Adrien Lavoillotte}, \bibinfo{person}{Makoto Miyazaki},
  {and} \bibinfo{person}{Lynn Heidmann}.} \bibinfo{year}{2020}\natexlab{}.
\newblock \bibinfo{booktitle}{\emph{Introducing {MLOps}}}.
\newblock \bibinfo{publisher}{O'Reilly Media}.
\newblock


\end{thebibliography}

\end{document}